\documentstyle[psfig]{elsart}
\newcommand{\be}{\begin{eqnarray}}
\newcommand{\ee}{\end{eqnarray}}
\newcommand\la{\langle}
\newcommand\ra{\rangle}
\begin{document}
\begin{frontmatter}
\title{Physical Complexity of Symbolic Sequences}
\author{C. Adami} and 
\author{N.J. Cerf$^1$}
\address{W. K. Kellogg Radiation Laboratory, 
California Institute of Technology,
Pasadena, California 91125, USA}
\thanks{Present address: Ecole
  Polytechnique, CP 165/56,  Universit\'e Libre
  de Bruxelles, B-1050 Bruxelles, Belgium}

\date{March 1999}

\maketitle
\begin{abstract}
  A practical measure for the complexity of sequences of symbols
  (``strings'') is introduced that is rooted in automata theory but
  avoids the problems of Kolmogorov-Chaitin complexity. This physical
  complexity can be estimated for {\em ensembles} of sequences, for
  which it reverts to the difference between the maximal entropy of
  the ensemble and the actual entropy given the specific environment
  within which the sequence is to be interpreted. Thus, the physical
  complexity measures the amount of information about the environment that is
  coded in the sequence, and is conditional on such an environment.
  In practice, an estimate of the complexity of
  a string can be obtained by counting the number of loci per string
  that are fixed in the ensemble, while the volatile positions
  represent, again with respect to the environment, randomness. We
  apply this measure to tRNA sequence data.

\end{abstract}
\end{frontmatter}
%
\section{Introduction}\label{intro}
The study of ``complex systems'', or more generally the science of
``complexity'', has enjoyed tremendous growth in the last decade,
despite the fact that complexity itself is only vaguely defined, and
many alternatives have been proposed over the years (see, e.g.,
\cite{KOL,CHA,BEN2,SLO2,bib_crutch}). In this paper we focus on the
complexity of {\em symbolic sequences} (mostly binary), as most
systems whose complexity we would like to estimate can be reduced to
them.

In searching for an adequate measure for the complexity of binary
strings, two limiting cases must be considered: the regular strings
(such as a sequence of only zeros) and the random ones. A good measure
of physical complexity is expected to yield a vanishing complexity for
both cases, while the ``intermediate'' strings that appear to encode a
lot of information are thought to be complex. Surprisingly, such a
measure has been difficult to define consistently. Here we propose a
measure of physical complexity that has the above-mentioned property,
but can consistently be defined within automata theory {\em and}
information theory.  Contrary to the intuition that the {\em
  regularity} of a string is in any way connected to its complexity
(as in Kolmogorov-Chaitin theory), we will argue here that such a
classification is, in the absence of an environment within which the
string is to be interpreted, quite meaningless. Rather, a ``regular''
string given by a uniform sequence can be made to represent anything
(e.g., all of Shakespeare's ``Hamlet'') if the coding is complicated
enough. In such a case, the coding rules represent part of the
sequence's environment.  Indeed, we propose that the complexity of a
string should, rather than focusing on its regularity, be determined
by analyzing its {\em correlation} with a physical environment.
Similarly, ``randomness'' is a meaningless concept without reference
to this environment. In general, we will find that a sequence can be
random with respect to one environment while perfectly ``meaningful''
with respect to another. In all cases, however, estimating the
complexity requires an ensemble of sequences and an environment with
which it is correlated.

In the next section we briefly review Kolmogorov-Chaitin complexity to
establish our notation and point out its well-known shortcomings as a
measure of physical complexity. Physical complexity is introduced in
Section 3, and its relation to notions from conventional (Shannon)
information theory is pointed out in Section 4. Practical
considerations for the estimation of complexities are presented in the
subsequent section. A real-life example is offered in Section 6, by
giving an estimate of the complexity of a tRNA sequence. Section 7
contains some speculations about the evolution of complexity in simple
living systems.

\section{Kolmogorov Complexity}

Kolmogorov-Chaitin (KC) complexity~\cite{KOL,CHA} is rooted in
automata theory~\cite{TUR}, and provides a measure for the {\em
  regularity} of a symbolic string.  Roughly, a string is said to be
``regular'' if the algorithm necessary to produce it on a universal
(Turing) automaton is shorter (with length measured in bits) than the
string itself. A simple example is a bit string with a repetitive
pattern, such as 1010101010... The minimal ``program'' enabling a
Turing automaton to write this string only requires the pattern 10,
the length of the string, and ``repeat, write'' instructions. A less
obvious example is the binary equivalent of, say, the first one hundred
digits of $\pi$. While random {\it prima facie}, a succinct
algorithm for a Turing machine can be written; as a consequence such a
string is classified as regular. Technically, the KC-complexity of a
string $s$ is defined (in the limit of long strings) as the length (in
bits) of the shortest program $p$ producing $s$ when run on universal
Turing machine $T$:
\begin{equation}
K(s) = {\rm min}\,\left\{ |p|:\,\, s = C_T(p)\right\}\;,
\label{kc-comp}
\end{equation}
where $|p|$ denotes the length of the program in bits and $C_T(p)$ is
the result of running program $p$ on Turing machine $T$. While
KC-complexity is only defined modulo the number of prefix instructions
(to be added to the program) necessary to simulate any other computer,
it becomes exact in the limit of infinite strings\footnote{A program
  executable on Turing machine $T$ can also be executed (with the same
  result) on any other universal computer $T^\prime$, provided that it
  is preceded by a {\em prefix} code. The relative difference in size
  of the minimal program on $T$ and $T^\prime$ due to the length of
  the prefix can be string dependent, but vanishes in the limit of
  infinite strings.}.

A simple consequence of (\ref{kc-comp}) is that algorithmically
regular strings have vanishing KC-complexity in the limit of infinite
strings, while ``random'' strings (such as binary strings obtained
from a coin-flip procedure) are assigned {\em maximum} KC-complexity,
i.e., for a random string $r$: \be K(r)\approx |r|\;. \label{random}
\ee Physically, and intuitively, this is unsatisfactory, and requires
us to rethink the very definition of ``random'' in a {\em physical}
world.  Our intuition demands that the complexity of a random string
ought to be zero, as it is somehow ``empty''. Furthermore, it has been
known for some time that randomness, from an automata-theoretic point
of view, must be {\em undecidable}, owing to G\"odel's undecidability
theorem applied to the halting problem: No halting computation can
possibly determine that a string is random, simply because such a
computation would render the string non-random~\cite{CHA75}.  In
Eq.~(\ref{random}), this problem is circumvented by allowing a random
string to be computed by a Turing machine if it is included {\it
  verbatim} on the program tape. Besides redefining the concept of
randomness, such a definition implies the (physically unsatisfactory)
property that random strings are maximally complex. Below, we show how
all these problems can be averted by insisting that {\em physical}
Turing machines never operate without context, i.e., without an
environment.

\section{Physical Complexity}
In order to define physical complexity, we first need to recall the
notion of ``conditional complexity'' defined earlier by Kolmogorov.
The idea implements precisely what we have called for earlier: that
the determination of the complexity of a sequence should depend---be
conditional on---the environment that the sequence is interpreted
within. The traditional Kolmogorov complexity, however, is only
conditional on the implicit rules of mathematics, and nothing else.
These rules are necessary to interpret the program on the tape, but
are usually not sufficient, as we shall see below.  Instead, let us
imagine a Turing machine that takes an infinite tape $e$ as {\em
  input} (which represent its {\em physical environment}) and that
includes the particular rules of mathematics of this ``world''.
Without such a tape, this Turing machine is incapable of computing
anything, except for writing to the output what it reads in the input.
Thus, in the absence of the infinite tape $e$ {\em all} strings have
maximal complexity. In other words, the aforementioned string that
represents $\pi$ {\em also} has maximal complexity if it is
unconditional on any rules (in contrast with the KC construction).

In this spirit, we can define the conditional complexity
$K(s|u)$~\cite{KOL,ZUR} as the length of the smallest program that
computes $s$ {\em from} $e$:
\begin{equation}
  K(s|e)=\min\ \{|p|:s=C_T(p,e)\}\;,
\end{equation} 
where $C_T(p,e)$ denotes the result of running program $p$ on Turing
machine $T$ given input string $e$. This is not yet a physical
complexity. Rather, the smallest program that computes $s$ from $e$,
in the limit of infinite strings, will contain bits that are entirely
{\em unrelated} to $e$, since, if they were not, they could be
obtained from $e$ with a program of size tending to zero. Thus,
$K(s|e)$ represents those bits in $s$ that are random with respect to
$e$. If $e$ were to represent the usual rules of mathematics only, the
complexity of $\pi$ (for example) {\em conditional} on $e$ reduces to the KC
complexity of $\pi$, i.e., zero.

The physical complexity can now be defined as the number of bits that
are meaningful in $s$ (that can be obtained from $e$ with a program of
vanishing size) and is given by the ``mutual
complexity''~\cite{KOL,ZUR}
\begin{equation}
K(s:e)=K_0(s)-K(s|e)\;.\label{mutcomp}
\end{equation}
Here, we have introduced the {\em unconditional} complexity $K_0(s)$,
i.e., the complexity given an empty input tape $e\equiv\emptyset$.
This is different from the Kolmogorov complexity $K(s)$ described above
because, in Kolmogorov's construction, the rules of mathematics were
given to the automaton.  As argued for above, {\em every} string is
random if no $e$ is specified, as non-randomness can only exist with
respect to a specific world, or environment. Thus, $K_0(s)$ is always
maximal, given by the length of $s$:
\begin{equation}
K_0(s)=|s|\;.  
\end{equation}
Consequently, as Eq.~(\ref{mutcomp}) represents the length of the
string minus those bits that can {\em not} be obtained from $e$,
$K(s:e)$ represents the number of bits that {\em can} be obtained, by
a computation with {\em vanishing} program size, from $e$. Thus, this
represents the physical complexity of $s$.  Let us investigate the
connection of these results to standard Shannon information
theory~\cite{bib_shannon}.

\section{Physical Complexity and Information Theory}
In fact, a moment's reflection reveals that $K(s:e)$, the complexity
of string $s$ given a description of the environment $e$, is not
practical, meaning that it can not, in general, be determined by
inspection. In other words, it is impossible to determine which, and
how many, of the bits of string $s$ correspond to information about
$e$. The reason is that in general, we are unaware of the {\em coding}
used to code information about $e$ in $s$, and as a consequence coding
and non-coding bits look entirely alike.  However, it is possible to
determine coding versus non-coding bits if we are given {\em multiple}
copies of a symbolic sequence that have adapted independently to the
environment within which it is to be interpreted, or more generally,
if a statistical {\em ensemble} of strings is available to us. In that
case, coding bits are revealed by non-uniform probability
distributions across the ensemble (``conserved sites''), whereas
random bits sport uniform distributions (``volatile sites'').  The
determination of complexity then becomes an exercise in information
theory.  Indeed, the link between automata theory and information
theory has been pointed out quite early, as it was realized~\cite{ZL}
that the average complexity $\la K\ra$, in the limit of infinitely
long strings tends to the entropy of the ensemble of strings
$S$\footnote{This holds for near-optimal coding. For strings $s$ that
  do not code perfectly we have $\la K\ra\geq H$ (see,
  e.g.,~\cite{ZUR90}.)}
\begin{equation}
\la K(s)\ra_S =\sum_s p(s) K(s)\approx H(S)=-\sum_s p(s)\log p(s)\;,  
\end{equation}
where string $s$ appears in the ensemble $S$ with probability $p(s)$.
Note that this is consistent with our determination that $K(s)$, in
the absence of an environment $e$, must equal the string's length.
Indeed, if nothing is known about the environment that the strings $s$
pertain to, the probability distribution $p(s)$ must be uniform
(principle of insufficient reason). As a consequence (if logarithms
are taken to base 2), $H(S)=|s|$, where $|...|$ as before denotes the
size of a string.  On the other hand, if an environment $e$ is given
we have some information about the system, and the probability
distribution is non-uniform. Indeed, it can be shown that for every
probability distribution $p(s|e)$ to find $s$ {\em given} $e$, we have
\begin{equation}
 H(S|e)\leq H(S)=|s|\;.  
\end{equation} 
as a result of the concavity of Shannon entropy.  The difference
between the maximal entropy $H(S)=|s|$ and $H(S|e)$, according to the
construction outlined above, should then represent the average number
of bits in strings $s$ taken from the ensemble $S$ that can be
obtained by zero-length universal programs from $e$:
\begin{equation}
\la K(s:e)\ra_S = \sum_sp(s)\,K(s:e)\approx H(S)-H(S|e) 
\equiv I(S:e)\;.\label{condinfo}
\end{equation}
In Eq.~(\ref{condinfo}), we used the usual definition of information
theory that the difference between the ``marginal'' entropy $H(S)$ and
the entropy of $S$ given $e$, $H(S|e)$, is just the information about
$e$ contained in the ensemble $S$. Note that strictly speaking,
$I(S:e)$ is not an information. Rather, an information is
obtained only if $I(S:e)$ is averaged over possible occurrences of $e$
in an ensemble $E$
\begin{equation}
I(S:E)=\sum_e p(e)\, I(S:e)\;.
\end{equation}
Despite this, we shall in the following continue to refer to $I(S:e)$
as the information about $e$ stored in $S$. We now ask the question
whether the physical complexity $I(S:e)$ is a measurable quantity.

\section{Estimating Entropies in Finite Ensembles}
\begin{figure}[t]
\caption{Entropy diagram for a string $s=ab$ (with subsequences $a$ and
  $b$) in environment
  $e$, with notations. The shaded region is the complexity 
  $I(ab:e)$. The areas that are not filled in in this diagram are irrelevant
  for the determination of complexity. \label{figtriple}}
\vskip 0.5cm
\centerline{\psfig{figure=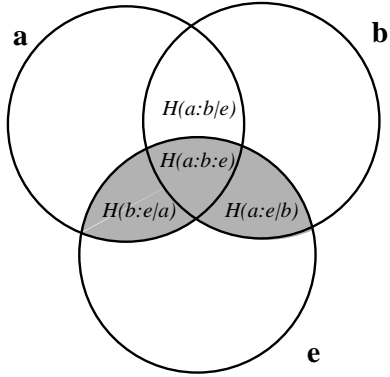,width=2.0in,angle=0}}
\end{figure}

In general, the entropy 
\begin{equation}
 H(S|e)=-\sum_s p(s|e)\log p(s|e)  
\end{equation}
can be estimated by sampling the probability distribution $p(s|e)$. In
a population of $N$ strings in environment $e$, the quantity $n(s)/N$
where $n(s)$ denotes the number of strings of type $s$, approximates
$p(s|e)$ arbitrarily well as $N\to\infty$. However, if the number of
different strings is very large as is typically the case for symbolic
strings (in particular genetic ones), the sampling error incurred from
a population that is not exponentially large can be overwhelming.
Indeed, it is known~\cite{BASH} that for symbolic strings that can
take on $M$ states, the sampling error in the entropy, to first order
in $1/N$, is
\begin{equation}
 \Delta H = \frac{M -1}{2N}\;, \label{error}
\end{equation}
if we agree to take logarithms to the base of the alphabet-size.
Thus, for strings of length $\ell$ constructed from an alphabet of
size $D$ only populations of the order $N\simeq D^\ell$ will ensure that the
finite-size error of the entropy is of order 1. In most practical
cases, such ensembles are unrealistic. Still, we may attempt to
estimate the entropy by summing the {\em per-site} entropies of the
string. Random sites, identified by a nearly uniform probability
distribution, contribute positively to the entropy whereas non-random
sites (which have strongly peaked distributions) contribute very
little.  Thus,
\begin{equation}
H(S|e)\approx \sum_i H(x_i|e)\;. \label{sumpersite}
\end{equation}
Using this estimate, however, introduces a systematic error which is
due to the fact that the particular code used to encode the sequence
may not be optimal. Let us consider as an example a sequence of length
two with positions $a$ and $b$, or more generally a sequence $s$
arbitrarily partitioned into subsequences $a$ and $b$: $s=ab$ (see
Fig.~\ref{figtriple}.)
The physical complexity in an environment $e$ is, according to
Eq.~(\ref{condinfo}), given by
\begin{equation}
I(S:e)=H(ab)-H(ab|e)\;,
\end{equation}
i.e., the entropy of string $s$ irrespective of any environment (or,
equivalently, averaged over all possible environments) minus the
entropy given the particular environment (see shaded region in
Fig.~1). This can be rewritten as
\begin{eqnarray}
I(S:e)&=&H(a)+H(b)-H(a|e)-H(b|e)-H(a:b:e)\\
&=&|s|-\sum_i H(x_i|e)-H(a:b:e)\;,
\end{eqnarray}
where we introduced the short hand $|s|=H(a)+H(b)$, i.e., the sum of
the unconditional per-site entropies is just the length of the
sequence, and $\sum_i H(x_i|e)$ is the sum of the per-site conditional
entropies. It is the latter which can be measured in actual populations.
Approximating the physical complexity by
\begin{equation}
C\simeq |s|-\sum_iH(x_i|e) \label{complex}
\end{equation}
[i.e., replacing the true entropy as in Eq.~(\ref{sumpersite})] thus
misses a piece $H(a:b:e)$, the center of the diagram in Fig.~1.  Now,
for perfect codes $H(a:b:e)$ actually vanishes, because all bits of
$s$ must be independent of each other when ignoring the environment
(i.e., the particular coding rules) $e$.  This implies that the
information is optimally stored in $s$ (it is perfectly compressed),
which is the case when taking the average of mutual KC complexity [see
Eq.~(\ref{condinfo})], as the limit of perfect codes is always implied
there. In physical ensembles, the error remains, and can only be
minimized by reducing the influence of correlations in $\sum_i
H(x_i|e)$. How this can be done approximately will be shown below
using tRNA as an example.  Note that $H(a:b:e)$ can be positive as
well as negative, implying that $C$ is neither a lower nor an upper
bound on the complexity.

\section{Complexity of Genomes}
As an example, let us find an approximation for the complexity of
biological sequences.  Our purpose here is to outline the practical
aspects of such a determination based on the complexity measure
proposed, rather than an examination of the feasibility of this method
with current technology. Consider for the purpose of illustration the
molecule tRNA, which consists out of 76 nucleotides that contort into
the well-known clover-leaf secondary structure (see Fig.~2), and whose
tertiary structure is essential for the translation of codons to amino
acids. If the complexity of this molecule is represented by the
genomic complexity of its sequence $s_{\rm RNA}$, then the complexity
of tRNA can be obtained by identifying the shortest description of the
random part of $s_{\rm RNA}$ (as this represents $K(S|u)$) and
subtracting it from the length of the molecule. Replacing $K$ by its
average, the physical complexity is then the length of the string
minus the remaining randomness according to Eq.~(\ref{complex}), and
should represent the {\em essence} which generates the RNA's function.
\begin{figure}[t]
\caption{Secondary structure of tRNA with 76 common positions, of
  which 52 are independent and thus useful in the determination of the
  sequence complexity. Fixed positions are black, moderately diverged
  ones are grey, and highly volatile ones are colored white 
  (from~\protect\cite{EIG}.)}
\vskip 0.5cm
\centerline{\psfig{figure=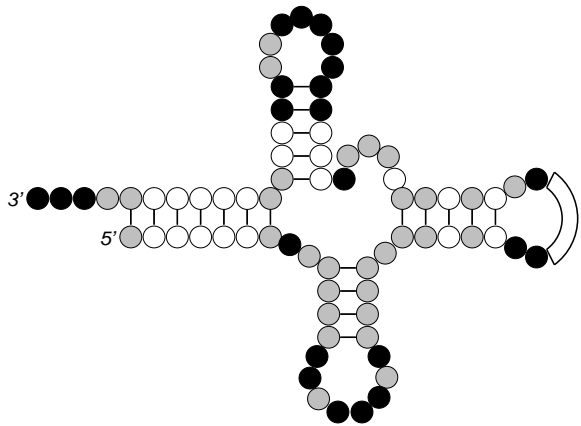,width=3.0in,angle=-90}}
\end{figure}

For tRNA from a given species, access to an ensemble of sequences
allows a classification of each position according to its volatility.
For {\em Bacillus subtilis} for example, we can use a sample of 32
aligned (structurally similar) sequences\footnote{Sequence data was
  obtained from the EMBL nucleotide sequence library~\cite{SSHS96}.}
to determine whether nucleotide positions are volatile (white in
Fig.~1), moderately diverged (grey) or fixed (black)~\cite{EIG}.
Counting the black (and to some extent the grey) sites should
approximate the complexity of the sequence.  But is this code optimal
for tRNA, or is there a substantial piece of the type $H(a:b:e)$
mentioned above? Such a piece represents correlations between sites in
the string which are {\em also} correlated to the environment, i.e.,
they are important for function. In fact, all those sites that form
Watson-Crick pairs in the secondary structure of the molecule (and
which show some variation) contribute to such an error, as their
contribution to the entropy in Eq.~(\ref{complex}) will be {\em
  double-counted}.  

An example of this is given in Fig.~3, which
depicts the entropy diagram of positions 27 and 31 in the sequence in
Fig.~2 (nucleotide 27 is paired with nucleotide 31 in the anticodon
stem.)
\begin{figure}[t]
\caption{Entropies for nucleotide positions 27 and 31 in the anticodon
  stem. The error in the entropy of the pair (27,31) contributing to
  the error in the complexity $I(S:e)$ is given by the center of the
  diagram. Not counting one of the two in the sum of per-site
  entropies eliminates this error.}
\vskip 0.5cm
\centerline{\psfig{figure=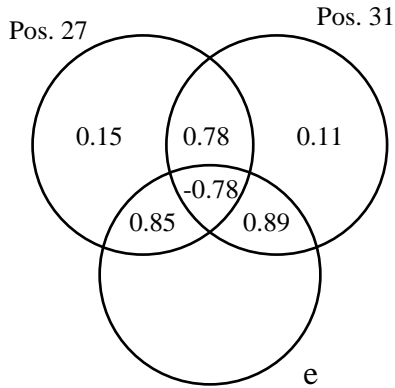,width=2.0in,angle=-90}}
\end{figure}
Counting paired positions only once (the per-site entropy of
base-paired positions are usually equal) and ignoring the anti-codon
leaves 52 ``reference positions''. Using the sequence data to
calculate $H(x_i)$ for each of the 52 positions reveals that the sum
of their entropies is approximately 29 nucleotides (58 bits.)  The
complexity estimate for this sequence is thus (ignoring the anti-codon
in $\ell$ also)
\begin{equation}
C= \ell-\sum_i \!{}^\prime\, H(x_i|e)=73-29=44\   {\rm nucleotides} \;,
\end{equation}
where $\sum^\prime$ denotes the sum over reference positions only. 
Naturally, only correlations due to base-pairing are eliminated in
this manner. Correlations due to other epistatic effects remain.
Note also that the estimate of the entropy $H\approx29$ is subject to the
finite sample error (\ref{error}), and is thus only accurate to 5\%. 

\section{Evolution of Complexity}
We can also extend our horizon and ask whether the evolution of
physical complexity displays the trend of evolution toward higher
complexity that seems evident in living systems (see,
e.g.,~\cite{BL}).  While we have an intuitive feeling that such an
evolution towards higher complexity is responsible for the emergence
of higher and higher organisms throughout time, such a statement must
be questionable as long as there is no unambiguous measure of
complexity.

Evolution of complexity can be observed explicitly in {\em artificial} living
systems~\cite{INTAL} which involve segments of (computer)-code
self-replicating in a noisy environment replete with information. In
such systems, an information landscape is specified by the user, and a
population of self-replicating computer programs is allowed to adapt
to it without external interference.  
More precisely, the ``accidental'' discovery (via random mutations) of
a sequence that benefits the string is ``frozen'' in the genome owing
to the higher replication rate of its bearer. The replication of each
string is effected by executing its code on a virtual computer.  As
such, these strings are analogous to catalytically active RNA
sequences that serve as the templates of their own reproduction.

The information-bearing sections of the code become apparent in
equilibrated populations of self-replicating code as they are {\em
  fixed}, while the volatile positions provide for genomic diversity
without storage of information. Again, the determination of volatility
of a site is only possible {\em statistically}, i.e., by examining
ensembles of members of the same ``species''. Adaptive events (in
which the replication rate increases) decrease the number of volatile
instructions if the sequence length stays constant, while a size
increase without commensurate acquisition of information increases
that number~\cite{INTAL}. Consequently, physical complexity (measured
sufficiently far away from an adaptive event in order to allow
equilibration) only increases in evolution.

According to the above arguments, the number of non-volatile
instructions in a code within a given environment represents an
estimate of the physical complexity of a particular species of string.
Placed in a different environment, the strings are meaningless; they
will not replicate anywhere except for the specific (real or virtual)
world they have evolved in. Furthermore, in a different world all
(previously) fixed positions will, under the influence of noise,
revert to volatile ones.  Thus, as emphasized throughout this paper,
the information content, or complexity, of a genomic string {\em by
  itself} (without referring to an environment) is a meaningless
concept.  In artificial living systems, the increase of physical
complexity, which coincides with increasing acquisition and storage of
information, can be monitored directly, and illustrates the usefulness
of this measure.  Note that this process of acquisition of information
constitutes, in the language of thermodynamics, to the operation of a
{\em natural} Maxwell-demon: the population performs random
measurements on its environment, and stores those ``results'' that
decrease the entropy, but rejects all others. Thus, the process can be
likened to a semi-permeable ``membrane'' for information, and the
physical complexity increases as a function of evolutionary time
(given a fixed environment) as the strings store more and more
information about that environment. Naturally, a change in environment
(catastrophic or otherwise) generally leads to a reduction in
complexity.  Such experiments suggest that physical complexity is
indeed the ``quantity that increases when self-organizing systems
organize themselves''~\cite{BEN3}.

This work was supported by the National Science Foundation
under Grant No. PHY-9723972. We are indebted to Tom Schneider for
pointing out Ref.~\cite{BASH}, and to C. Ofria and W.H. Zurek for
discussions.

\end{document}